\documentstyle[prb,aps,epsfig,floats,goodfloat]{revtex}

\newcommand{\bc}{\begin{center}}
\newcommand{\ec}{\end{center}}
\newcommand{\be}{\begin{equation}}
\newcommand{\ee}{\end{equation}}
\newcommand{\beqn}{\begin{eqnarray}}
\newcommand{\eeqn}{\end{eqnarray}}
\newcommand{\nsw}{N_{\mathrm{sweep}}}
\newcommand{\nsa}{N_{\mathrm{samp}}}
\newcommand{\varql}{{\mathrm{Var}(\ql)}}
\newcommand{\ql}{q_{l}}

\begin{document}
\draft

\twocolumn[\hsize\textwidth\columnwidth\hsize\csname@twocolumnfalse%
\endcsname

\title{
Monte Carlo Simulations of Spin Glasses at Low Temperatures
}

\author{Helmut G. Katzgraber, Matteo Palassini and A. P. Young}
\address{Department of Physics, University of California, Santa Cruz, 
CA 95064}

\date{\today}

\maketitle

\begin{abstract}
We report the results of Monte Carlo simulations on several spin glass models
at low temperatures. By using the parallel tempering (Exchange Monte Carlo)
technique we are able to
equilibrate down to low temperatures,
for moderate sizes, and hence the data should not be affected by critical
fluctuations. Our results for short range models are consistent with a picture
proposed earlier that there are large scale excitations which cost only a
finite energy in the thermodynamic limit, and these excitations have a surface
whose fractal dimension is less than the space dimension. For the infinite
range Viana-Bray model, our results obtained for a similar number of spins
are consistent with standard replica symmetry breaking.
\end{abstract}

]


\section{Introduction}
There has recently been a renewed interest in the nature of the spin glass
phase.  Two principal theories have been investigated: the
``droplet model'' proposed by Fisher and Huse\cite{fh},
(see also Refs.~\onlinecite{bm,mcmillan}),
and the replica symmetry
breaking (RSB) picture of Parisi\cite{parisi,mpv,bindery}.
These scenarios and some others have also been considered by Newman and
Stein\cite{ns}.
An important difference
between these models concerns the number of large-scale, low energy
excitations. RSB theory follows the exact solution of the
infinite range Sherrington-Kirkpatrick (SK) model in predicting that
there are excitations which involve turning over a
finite fraction of the spins and which cost only a {\em finite} amount of
energy in the thermodynamic limit.  In addition, the surface of these
large, finite energy excitations is expected\cite{qlink} to be
space filling which means that the fractal
dimension of their surface, $d_s$, is equal to the space dimension, $d$.
The droplet theory argues that the lowest energy excitation involving
a given spin and which has linear spatial extent $L$ typically costs an energy
$L^\theta$, where $\theta$ is a (positive) exponent. Hence, in the
thermodynamic limit, excitations which flip a finite fraction of the spins cost
an {\em infinite} amount of energy. The droplet theory also predicts that
$d_s < d$.

Recently,
Krzakala and Martin\cite{km} (KM),
and two of us\cite{py4} (PY),
have argued
that a straightforward interpretation of their numerical results at zero
temperature is intermediate between the droplet and RSB pictures in that there
appear to be large scale excitations whose energy does not increase with
size, but these have a surface with $d_s < d$.
This interpretation of the results of KM and PY has, however, been recently
challenged (though in opposite senses) by
Marinari and Parisi\cite{mp} and by Middleton\cite{aam}. There has also been
recent debate\cite{hg,mprz,pypmj,hed}
as to whether the $\pm J$ model has similar behavior to that of a model with a
continuous distribution. 
In the scenario of KM and PY
it is necessary to introduce
{\em two} exponents which describe the growth of the energy of an excitation
of scale $L$: (i) $\theta \ (> 0)$ such that $L^\theta$ is the typical change
in energy when the boundary conditions are changed, for example from periodic
to anti-periodic, and (ii) $\theta^\prime$, which characterizes the energy of
clusters excited within the system for a {\em fixed} set of boundary conditions.
In this paper we test whether the picture proposed by KM and PY
is compatible with finite temperature Monte
Carlo simulations. 

Several previous Monte Carlo simulations\cite{rby,marinari,zuliani,qlink} 
have found evidence
for finite energy large scale excitations by looking at the order parameter
function $P(q)$. In the thermodynamic limit this has delta functions at (plus
or minus) the
Edwards-Anderson order parameter, $q_{EA}$, corresponding to ordering within a
single valley, and, according to RSB theory, a tail with a finite weight
extending down to $q=0$. In the droplet theory, $P(q)$ is trivial, i.e. has
only delta functions at $\pm q_{EA}$, though in a finite system there is a
weight at the origin
which vanishes with increasing $L$ like\cite{fh} $L^{-\theta}$.

These earlier Monte Carlo studies have found that the weight of $P(q)$ at 
the origin is independent of the systems size for temperatures down to 
$T\simeq 0.7\,T_c$ in
three dimensions\cite{marinari} and $T\simeq 0.6\, T_c$ in 
four dimensions \cite{zuliani}. 
However, these studies have been criticized\cite{moore1,moore2}
as being too close to the critical point, so that the results are affected by
critical fluctuations and very much larger sizes would be needed at these
temperatures to see the asymptotic behavior of the low-temperature
spin-glass state\cite{comment1}.
Refs.~\onlinecite{moore1,moore2} also argue, however, that
clear evidence for droplet
theory behavior could be seen even for quite small sizes at {\em
very}\/ low temperatures.

In this paper we check this prediction 
by performing Monte
Carlo simulations in the low temperature region, though with an
admittedly modest range of sizes, using the ``parallel tempering'' Monte Carlo
method\cite{huk_nem,marinari_buda}, also known as Exchange Monte Carlo.
One difficulty with this approach is to
ensure equilibration since the technique
proposed earlier by one of us and Bhatt\cite{by} for conventional Monte Carlo
does not work for parallel tempering.
Here we use an alternative method, valid for the important case of
a Gaussian distribution (which we use here), and which is closely related to
the approach of Ref.~\onlinecite{huku_sk} for the SK
model.

Both in three and four dimensions, we find a tail in $P(q)$
which is independent of size (up to the sizes studied), for
temperatures down to 
$T \simeq 0.2 \,T_c$ in 3D and $T\simeq 0.1 \,T_c$ in 4D,
in contrast to the
prediction of Refs.~\onlinecite{moore1,moore2}.
We also find that data for the ``link overlap'',
defined below, fits well a description with $d_s < d$, though the
extrapolation to the thermodynamic limit is quite large here. Thus our results
are completely consistent with the earlier proposal of KM and PY.

We consider the short-range Ising spin glass in three and four
dimensions, and, in addition, the Viana-Bray\cite{vb}
model. The latter is
infinite range but with a finite average
coordination number $z$, and is expected to show RSB behavior.
All these models have a finite
transition temperature. In the 3D case, the exponent $\theta$ obtained from
the magnitude of the change of the ground state energy when the boundary
conditions are changed from periodic to anti-periodic is\cite{theta-3d}
about $0.2$, whereas in 4D it is much larger\cite{theta-4d},
about 0.7.

The Hamiltonian is given by
\begin{equation}
{\cal H} = -\sum_{\langle i,j \rangle} J_{ij} S_i S_j ,
\label{ham}
\end{equation}
where, for the short range case, the sites $i$ lie on a 
simple cubic lattice in dimension $d=3$ or 4 with $N=L^d$ sites 
($L \le 8$ in 3D, $L \le 5$ in 4D), $S_i=\pm
1$, and the $J_{ij}$ are nearest-neighbor interactions chosen according to a
Gaussian distribution with zero mean and standard deviation unity. Periodic
boundary conditions are applied. 
For the Viana-Bray
model each spin is connected with $z=6$ other
spins on average chosen randomly (but
with the constraint that the total number of bonds is {\em exactly}\/ $3N$).
We allowed the local coordination to fluctuate 
which is different from the
more familiar Viana-Bray model in which each
site has exactly the same coordination number,
but we expect the properties of
the two models to be very similar.
The width of the Gaussian distribution is again unity, and
the range of sizes is $N \le 700$.

Our attention will focus primarily on two quantities:
the spin overlap, $q$,
defined by
\begin{equation}
q = {1 \over N} \sum_{i=1}^N S_i^{(1)} S_i^{(2)} ,
\label{q}
\end{equation}
where ``$(1)$'' and ``$(2)$'' refer to two copies (replicas)
 of the system with identical
bonds, and the link overlap, $\ql$, defined by
\begin{equation}
\ql = {1 \over N_b} \sum_{\langle i, j \rangle}
S_i^{(1)} S_j^{(1)} 
S_i^{(2)} S_j^{(2)} .
\label{ql}
\end{equation}
In the last equation, $N_b$ is the number of bonds ($Nz/2$ for the models
considered here, where $z$ is the coordination number),
and the sum is over all pairs of spins $i$ and $j$ which are
connected by bonds. 
The advantage of calculating $\ql$ as well as $q$ is that
if two spin configurations differ by flipping a large cluster then $q$
differs from unity by an amount proportional to the {\em volume}\/ of the
cluster while $\ql$ differs from unity by an amount proportional to the {\em
surface}\/ of the cluster.

\section{Equilibration}
Simulations of spin glasses at low temperatures are now possible, at
least for modest sizes, using the parallel
tempering Monte Carlo method\cite{huk_nem,marinari_buda}.
In this technique, one simulates several identical replicas of the system
at different temperatures, and, in addition to the usual local moves, one
performs global moves in which the temperatures of two replicas (with
adjacent temperatures) are exchanged. It turns out to be straightforward to
design an algorithm which satisfies the detailed balance condition, and it will
also have a good acceptance ratio if the temperatures are fairly close
together.  In this way, the temperature of a given replica wanders up and
down in a random manner, and each time the temperature goes low the system is
likely to end up in a different valley of the energy landscape. Thus different
valleys are sampled in much less time than it would take for the system to
fluctuate between valleys if the temperature stayed fixed.

We choose a set of temperatures $T_i, i = 1, 2, \cdots , N_T$, in order that
the acceptance ratio for the global moves is satisfactory, typically greater
than about $0.3$. Since, at each temperature, we need two copies of the system
to calculate $q$ and $\ql$ as shown in Eqs.~(\ref{q}) and (\ref{ql}), we
actually run 2 sets of $N_T$ replicas and perform the global moves
independently in each of these two sets.

To believe
results of simulations carried out at low temperatures
it is essential to
have a sound criterion for equilibration. The technique pioneered by 
Ref.~\onlinecite{by}
does not work with parallel tempering Monte Carlo because the
temperature does not stay constant.
However, another method can be used
for a Gaussian distribution of exchange interactions, which is 
a common and convenient choice.
It depends on an identity first noted a long time ago
by Bray and Moore~\cite{bmq2} for the SK model. Here we give the
corresponding result for the short range case.
We start with the expression for the average energy per site,
\begin{equation}
U = - {1\over N} \sum_{\langle i,j \rangle} [\, J_{ij} \langle S_i S_j
\rangle_{_T} \, ]_{av} ,
\end{equation}
where $\langle \cdots \rangle_T$ denotes
the Monte Carlo average for a given set
of bonds, and $[ \cdots ]_{av}$ denotes an average over the (Gaussian) bonds
$J_{ij}$. 
One can perform an integration by parts over the $J_{ij}$ to relate $U$ to
the average link overlap defined in Eq.~(\ref{ql}), i.e.
\begin{equation}
\langle \ql \rangle \equiv  {1 \over N_b}
\sum_{\langle i, j \rangle}
[ \langle S_i S_j \rangle_{_T}^2 ]_{av} = 1 - {T |U| \over (z/2) J^2} ,
\label{equil_cond}
\end{equation}
where the brackets $\langle \cdots \rangle$ indicate both a
Monte Carlo average and 
and an average over disorder,
$J^2$ is the variance of the interactions (set equal to unity in this
paper),
the sum is over sites $i$ and $j$ connected by bonds (each pair
counted once), and
the factor of $z/2$ arises because there
are $z/2$ times as many bonds as sites. A very similar approach 
has been used to test equilibration of the parallel tempering method for
the SK
model\cite{huku_sk}, except that in that case the square of the spin glass
order parameter appears rather than $\ql$.

We start the simulation by randomly choosing
the spins in the $2 N_T$ replicas
to be uncorrelated with each other. This means that
the two sides of Eq.~(\ref{equil_cond}) should
approach the equilibrium value
from {\em opposite}\/ directions for the following reason. 
The data for $\langle \ql \rangle$ will be too small if the system is not
equilibrated because the random start means that the spins are
initially further away from each
other in configuration space than they will be in
equilibrium, whereas initially the energy will not be as negative as in
equilibrium so the right hand side of Eq.~(\ref{equil_cond}) will initially be
too high. 
Hence we expect that if the two
sides of Eq.~(\ref{equil_cond}) agree then the system is in equilibrium.

\begin{figure}
\begin{center}
\epsfxsize=\columnwidth
\epsfbox{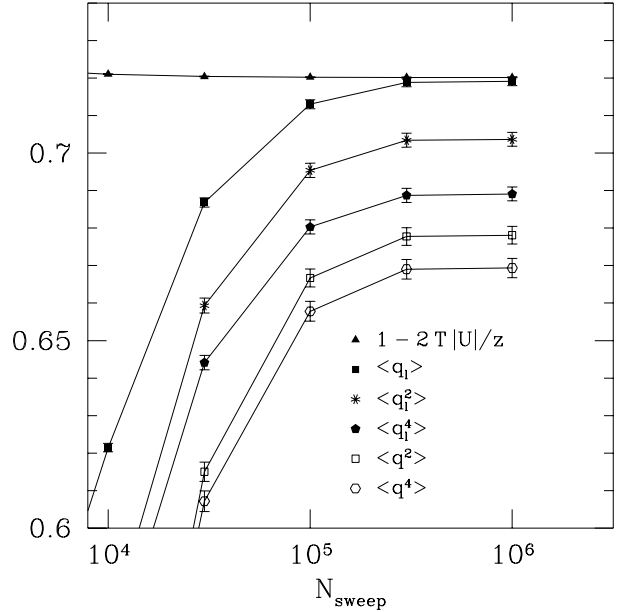}
\end{center}
\caption{
The solid squares are results for the average link overlap, defined by
Eq.~(\ref{ql}), as a function of the number of Monte Carlo sweeps, $\nsw$,
that each of the $2 N_T$ replicas performs.
Averaging was performed over the last half of the sweeps indicated.
The triangles are obtained from the energy in the
way indicated, and should agree
with the results for $\ql$ if the system is in equilibrium, as shown in
Eq.~(\ref{equil_cond}). The two sets of data approach each other from opposite
directions and then do not appear to change at larger number of sweeps,
indicating that they have equilibrated. 
We also show data for higher moments of $q$ and $\ql$. They appear to be
independent of the number of sweeps once the $\ql$ data has equilibrated. 
The data for the different moments has been shifted upwards
by the following amounts
for better viewing: $\langle q^2 \rangle$ by $0.11$, $\langle q^4 \rangle$ by
$0.48$, $\langle \ql^2 \rangle$ by $0.17$, $\langle \ql^4 \rangle$ by $0.375$.
These results are for $ d=3, T=0.5$ and $L=8$, and the data is averaged
over 3891 samples.
}
\label{equil}
\end{figure}

For illustration purposes we show in Fig.~\ref{equil}
how this works for three-dimensions
with $L=8$ and $T=0.5$ (to be compared with\cite{3d}
$T_c \simeq 0.95$).
The data for $\langle \ql \rangle$ increases as the length of the simulation
increases while that determined from the energy decreases (by a lesser
amount). Once the two agree they do not appear to change at longer times,
indicating that they have reached equilibrium, as expected. 
Furthermore, the data for different moments of the spin overlap $q$
as well as the link overlap $\ql$ appear to saturate
when $\ql$ has equilibrated so it does not appear that there are longer
relaxation times for $q$ than for $\ql$. We also checked that the whole
distribution $P(q)$ does not change with time once the two estimates for
$\langle \ql \rangle$ agree.
By presenting results for an intermediate temperature,
rather than the lowest temperature, in Fig.~\ref{equil},
we show that the results do not change for
times {\em longer}\/ than that needed for $\ql$ and $1-2T|U|/z$ to agree. The
length of the simulation was chosen so that, at
{\em lowest}\/ temperature, the data for $\ql$ and $1-2T|U|/z$ {\em just
converged}\/.

\section{Results}
\subsection{Three Dimensions}
In Table~\ref{3d-tab}, we show $\nsa$, the
number of samples, $\nsw$, the total number of sweeps performed by each
set of spins (replicas), and $N_T$,
the number of temperature values, used in the 3D simulations.
For each size, the largest temperature is 2.0 and the
lowest temperature is 0.1. However, for $L=8$, the data for the two lowest
temperatures, $T=0.10$ and 0.15, are not fully equilibrated so data at these
temperatures has been ignored and the lowest temperature used in the analysis
is 0.20. This is to be compared with\cite{3d}
$T_c \approx 0.95$. The set of temperatures is determined by requiring that 
the acceptance ratio for global moves is satisfactory for the largest size,
$L=8$, and for simplicity the same temperatures are also used for the smaller
sizes. For $L=3,4,5$, and 6, the acceptance ratio for global moves is 
greater than about 0.6 in average and is always greater than 0.3 for each pair
of temperatures. For $L=8$ the average acceptance ratio is 0.41, 
and the lowest value is 0.12.

\begin{figure}
\begin{center}
\epsfxsize=\columnwidth
\epsfbox{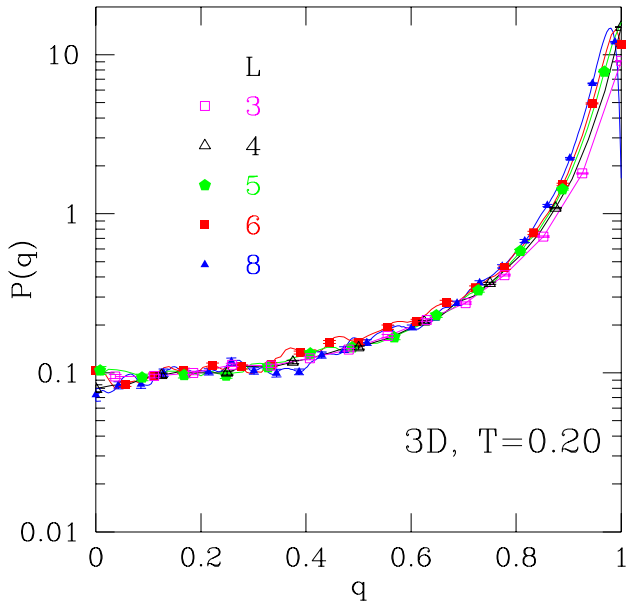}
\end{center}
\caption{
Data for the overlap distribution $P(q)$ in 3D at $T=0.20$. Note that the
vertical scale is logarithmic to better make visible both the peak at large
$q$ and the tail down to $q=0$. In this and other similar figures in the paper,
we only display {\em some}\/ of the data points
as symbols, for clarity, but the lines connect {\em all}\/ the data points.
This accounts for the curvature in some of the lines in between neighboring
symbols. In this paper all distributions are normalized so that the area shown
under the curve is unity.
}
\label{pq0.20_3d}
\end{figure}

\begin{figure}
\begin{center}
\epsfxsize=\columnwidth
\epsfbox{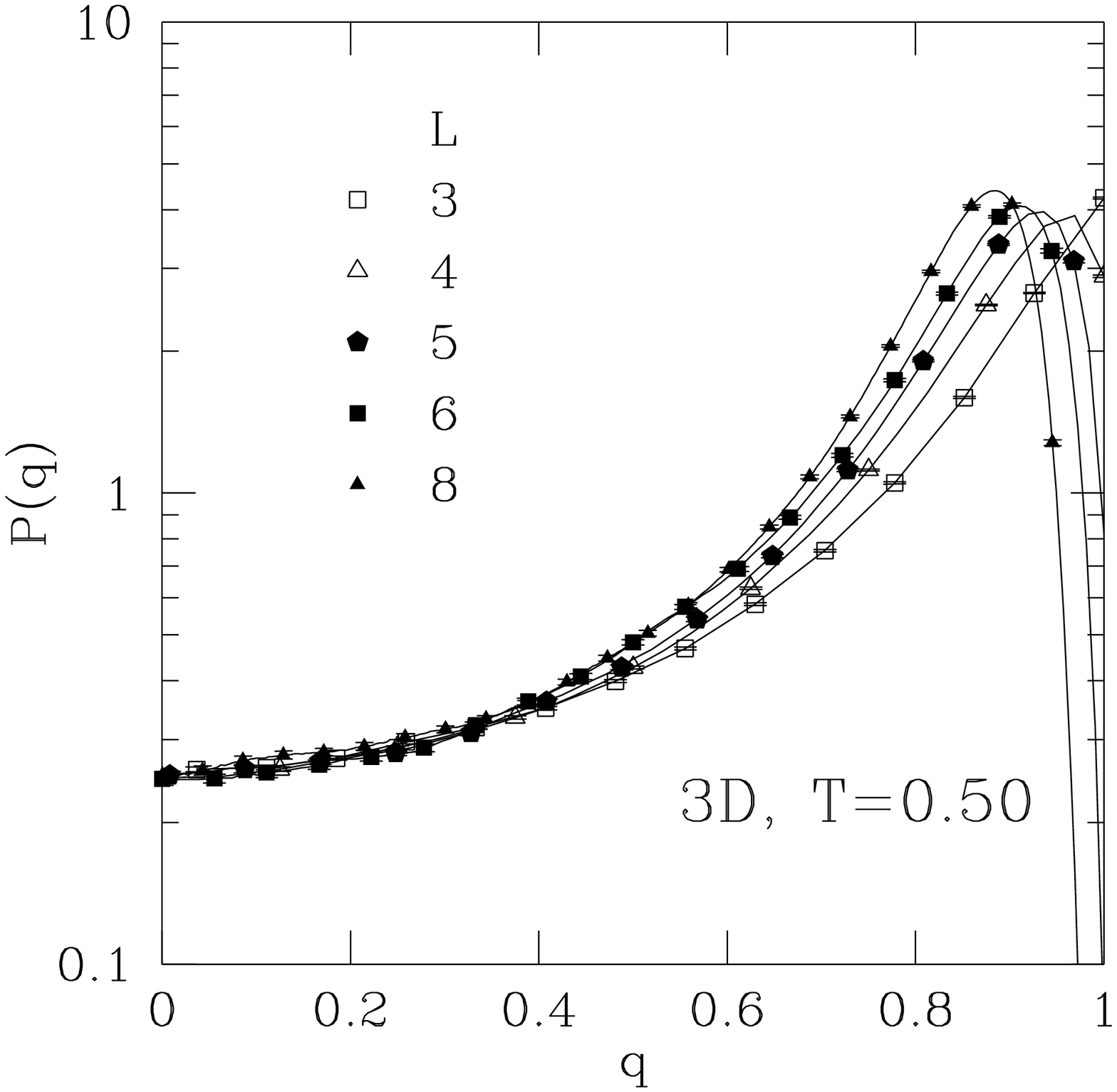}
\end{center}
\caption{
Same as for Fig.~\ref{pq0.20_3d} but at $T=0.50$.
}
\label{pq0.50_3d}
\end{figure}

\begin{table}
\begin{center}
\begin{tabular}{lrrr}
L  &  $\nsa$  & $\nsw$ & $N_T$  \\ 
\hline
3 & 15000  &     $10^4$  &   18          \\
4 & 16000  &     $10^4$  &   18          \\
5 &  7590  &     $10^5$  &   18          \\
6 &  4539  &   $3 \times 10^5$  &   18   \\
8 &  3891  &     $10^6$  &   18

\end{tabular}
\end{center}
\caption{Parameters of the simulations in three dimensions. $\nsa$ is the
number of samples (i.e. sets of bonds), $\nsw$ is the total number of sweeps
simulated for each of the $2 N_T$ replicas for a single sample,
and $N_T$ is the number of
temperatures used in the parallel tempering method.
}
\label{3d-tab}
\end{table}

\begin{table}
\begin{center}
\begin{tabular}{lrrr}
$L$  &  $\langle E \rangle$  & $\langle E_0\rangle$ & $N_0$  \\ 
\hline
4 &  $-106.49 \pm  0.05$  & $ -106.60 \pm  0.03$ & 50000          \\
6 &  $-364.06 \pm  0.17$  & $ -364.94 \pm  0.06$  &  39246   \\
8 &  $-867.04  \pm 0.27 $ & $ -868.20 \pm  0.15$  &   13302
\end{tabular}
\end{center}
\caption{Average energy $\langle E \rangle$ at $T=0.2$ 
and average ground state energy $\langle E_0 \rangle$, for several
sizes in three dimensions. $N_0$ is the number of samples used to compute the
average $\langle E_0 \rangle$ using a hybrid genetic algorithm.}
\label{tab_gs_3D}
\end{table}

In Table~\ref{tab_gs_3D} we compare the average total energy $\langle E \rangle
=N U$ at $T=0.20$ with
the average ground state energy obtained  by finding the ground 
state of each sample with a hybrid genetic algorithm, as discussed elsewhere\cite{py2}.
The two energies are very close together, indicating
that at this temperature our data are unlikely to be affected by the 
critical point.

\begin{figure}
\begin{center}
\epsfxsize=\columnwidth
\epsfbox{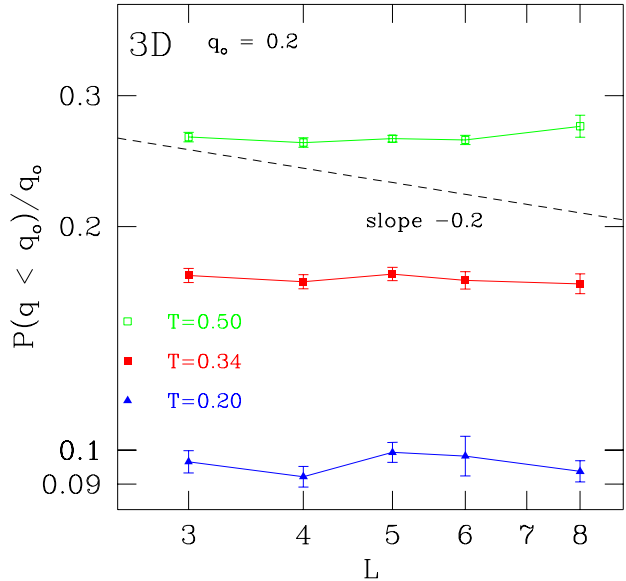}
\end{center}
\caption{
Log-log plot of $P(0)$, the spin overlap at $q=0$, against $L$ in 3D.
The data is independent of size within the error bars. The
dashed line has slope $-0.20$, which is the estimated value of $-\theta$.
Asymptotically, the data should be parallel to this line according to the
droplet theory. 
}
\label{p0_3d}
\end{figure}

Figs.~\ref{pq0.20_3d} and \ref{pq0.50_3d}
show (symmetrized) data for $P(q)$ at temperatures 0.20
and 0.50. There is clearly a peak for large $q$ and a tail down to $q=0$.
At both temperatures one sees that the tail in the distribution is
essentially independent of size.
A more precise determination of the size dependence of $P(0)$ is shown in
Fig.~\ref{p0_3d} where, to improve statistics, we average over the
(discrete) $q$-values with $|q| < q_\circ$, with $q_\circ=0.20$.
If general, we expect that $P(0) \sim L^{-\theta^\prime}$, 
where we allow $\theta^\prime$ to be different from $\theta$, the latter
being obtained from boundary condition changes.
In the droplet picture\cite{fh} $\theta^\prime=\theta$.
The dashed line in Fig.~\ref{p0_3d} has slope $-0.20$
corresponding to the estimated value\cite{theta-3d} of $-\theta$,
so in the droplet picture the data is expected to follow a track 
parallel to this line.  The actual size dependence is clearly 
much weaker than this, and consistent with a constant $P(0)$,
which implies that
the energy to create a large
excitation does not increase with size, and therefore $\theta^\prime \simeq 0$.

More precisely, a two-parameter fit of the data in Fig.~\ref{p0_3d} with
the form $a L^{-\theta^\prime}$,
gives $\theta^\prime=0.01 \pm 0.05$ for $T=0.20$, $\theta^\prime=0.02 \pm
0.02$ for $T=0.34$, and $\theta^\prime=-0.01 \pm 0.02$, for $T=0.50$. 
Different values of
$q_\circ$ give similar results for the fits.
We also tried a one-parameter fit in which $\theta^\prime$ is fixed.
Assuming $\theta^\prime = 0$ the goodness-of-fit parameter Q is
$0.34, 0.77$ and $0.57$ for $T=0.20, 0.34$ and $0.50$ respectively,
whereas assuming $\theta^\prime = 0.20$ the goodness-of-fit parameters
are very low,
$1.4 \times 10^{-4}, 2.8 \times 10^{-6}$ and $7.6 \times 10^{-15}$
respectively.
Hence, just considering statistical errors for the sizes studied, the data is
compatible with $\theta^\prime = 0$ and not with $\theta^\prime = 0.20$.

Refs.~\onlinecite{moore1,moore2} studied $P(0)$ by the Migdal-Kadanoff
approximation, which  is known to yield the droplet picture asymptotically.
They find that, although the behavior of $P(0)$
at higher temperatures is masked by critical point effects, data at low
temperatures, such as those considered here, {\em should}\/ show the droplet
behavior. That we find quite different results indicates that the
Migdal-Kadanoff approximation is not applicable to such small sizes. However,
our data still do not rule out the possibility that the droplet theory, or
some other theory, might be correct at larger sizes.

\begin{figure}
\begin{center}
\epsfxsize=\columnwidth
\epsfbox{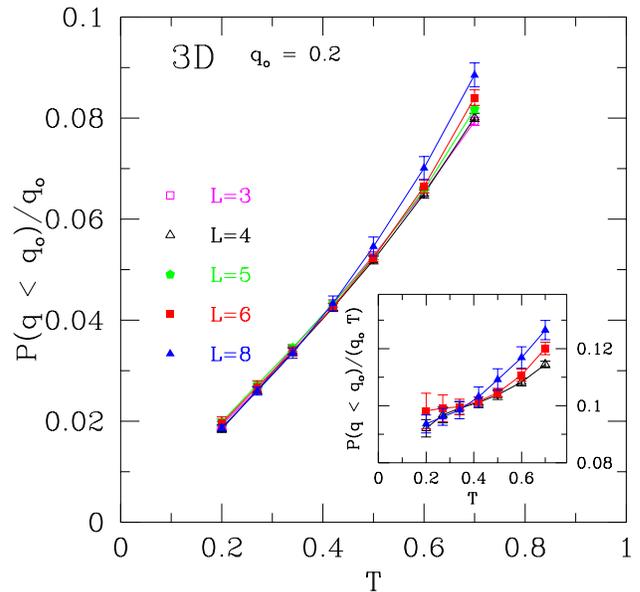}
\end{center}
\caption{
Data for $P(0)$ as a function of temperature in 3D for different values of $L$.
The inset shows $P(0)/T$ vs.~$T$.
}
\label{p0T_3d}
\end{figure}

In Fig.~\ref{p0T_3d} we show the data for $P(0)$ versus $T$. We see an
approximately linear
decrease of the data as $T \rightarrow 0$. Note though, that there
is some non-linearities as shown in the figure's inset.

\begin{figure}
\begin{center}
\epsfxsize=\columnwidth
\epsfbox{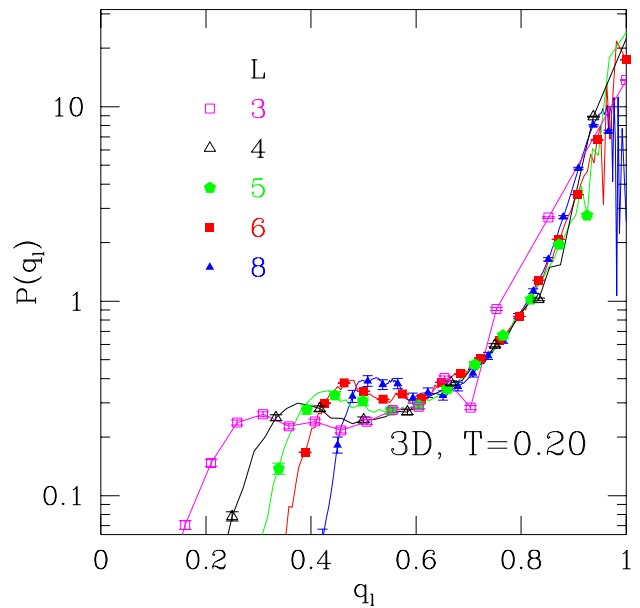}
\end{center}
\caption{
The distribution of the link overlap in 3D at $T=0.20$ for different sizes.
Note the logarithmic vertical scale.
}
\label{pqb0.20_3d}
\end{figure}

Fig.~\ref{pqb0.20_3d} shows the distribution of the link overlap
$\ql$ at $T=0.20$.
We see that there is a large peak at $\ql$ close to unity (with structure coming
from the allowed discrete values of $\ql$) and a much weaker peak (note the
logarithmic vertical scale) for smaller $\ql$ which grows slowly with increasing
$L$ and moves to larger values of $\ql$.
We will refer to
this feature again below when we discuss the Viana-Bray model.

\begin{figure}
\begin{center}
\epsfxsize=\columnwidth
\epsfbox{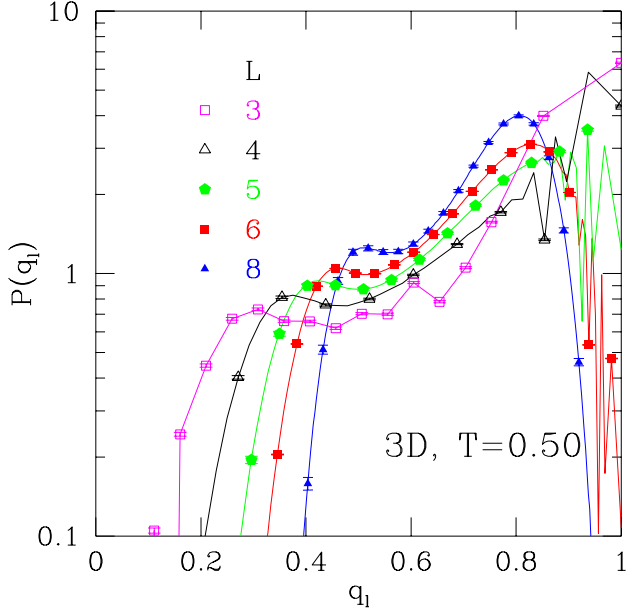}
\end{center}
\caption{
Same as for Fig.~\ref{pqb0.20_3d} but at $T = 0.50$.
}
\label{pqb0.50_3d}
\end{figure}

\begin{figure}
\begin{center}
\epsfxsize=\columnwidth
\epsfbox{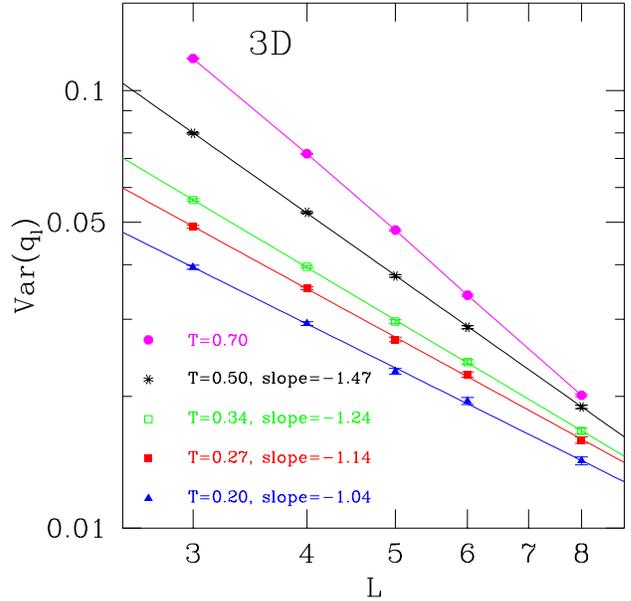}
\end{center}
\caption{
Log-log plot of the variance of $\ql$ as a function of size in 3D at several
temperatures. 
The data for $T = 0.50$ and $T = 0.70$ are multiplied by $1.2$ and $1.7$
respectively for better viewing.
The data for $T = 0.70$ is somewhat curved and so does not fit well a power
law.
}
\label{varqb_3d}
\end{figure}

The variance of $P(\ql)$ is shown in Fig.~\ref{varqb_3d}
for several low temperatures.
The data is consistent with a power law decrease to zero, i.e.
\begin{equation}
\varql \sim L^{-\mu_l} ,
\label{varqlpower}
\end{equation}
where the power $\mu_l$ seems to vary somewhat with $T$. This variation is
probably due to corrections to scaling coming from
the shift with $T$
of the complicated peak structure at large $q$ seen
in Fig.~\ref{pqb0.20_3d}.

The asymptotic value of $\mu_l$ is related to the
exponents $\theta^\prime$ and $d_s$ that have been mentioned
earlier\cite{dbmb}. To see this,
assume that the non-zero variance is largely
due to the excitation of a single large
cluster of size of order $L$. Its energy is of order
$L^{\theta^\prime}$, and the probability that thermal fluctuations can create
it is of order $T/L^{\theta^\prime}$, assuming a constant density of states
for these excitations.  One
minus the link overlap between the two states is of order
$L^{-(d-d_s)}$ because there is only a contribution to $1-\ql$
from the surface of the
cluster. Hence there is a probability $T L^{-\theta^\prime}$ of getting
a $\delta\ql$ of order $L^{-(d-d_s)}$, so the variance goes like
$T L^{-\mu_l}$ where
\begin{equation}
\mu_l = \theta^\prime + 2(d - d_s) .
\end{equation}

An extrapolation of our results for $\mu_l$ to $T=0$ gives
$0.76 \pm 0.03$. Assuming $\theta^\prime = 0$ this implies
\begin{equation}
d - d_s = 0.38 \pm 0.02
\end{equation}
which is consistent with the $T=0$
result of PY, namely $d - d_s = 0.42 \pm 0.02$.

\begin{figure}
\begin{center}
\epsfxsize=\columnwidth
\epsfbox{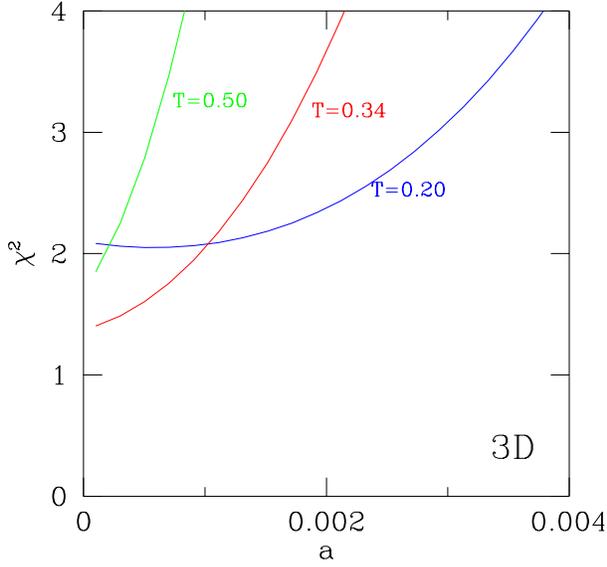}
\end{center}
\caption{
The $\chi^2$ of the fit in Eq.~(\ref{fit}) (in which $a$ is fixed and
$\ln b$ and
$c$ are fit parameters) for different values of $a$.  The results are for the 3D
model.
}
\label{chi2_3d}
\end{figure}

We have also looked at more general fits of the form
\begin{equation}
\varql = a + {b \over L^c} ,
\end{equation}
to see to what extent the data can rule out a non-zero value of $a$.
This is of interest because a non-zero value for $a$ is 
required by standard replica symmetry breaking theory. We carried out
fits of the form
\begin{equation}
\ln[\varql - a] = \ln b - c \ln L ,
\label{fit}
\end{equation}
in which $a$ is fixed, and $\ln b$ and $c$ are 
the fit parameters. The $\chi^2$ of the fit
is then determined as a function of $a$ and the results are shown in
Fig.~\ref{chi2_3d}.

One sees that the minimum of $\chi^2$ is for $a=0$ and the range of $a$ in
which $\chi^2$ has increased by less than unity relative to the $a=0$ value
is 
$a < 5.3 \times 10^{-4}$ for $T=0.50$ and $a < 1.3 \times 10^{-3}$ for
$T = 0.34$. The
width of the distribution of $\ql$ is $\sqrt{a}$ which has values 
$0.023$ and $0.036$ respectively. Thus while, our data cannot rule out
a non-zero
value for the width of the distribution of $\ql$ in the thermodynamic limit,
it does suggest that this value, if non-zero, must be very small.
We note, however, that a rather small value of $a$ is not
unreasonable in RSB. For example, if $P(q_l)$ consists of two delta
functions at a distance of 0.1, whose weights are 0.1 and
0.9 respectively, then the value of $a$ is 0.0009.

\subsection{Four Dimensions}
In four dimensions we present results down to a temperature of $0.20$,
compared with\cite{4d} $T_c \approx 1.80$. 
Parameters of the simulations are shown in 
Table~\ref{4d-tab}. For each size, the largest temperature is 2.80 and the
lowest is 0.20. The acceptance ratio for global moves is always greater than
about $0.5$ for $L=3$, about $0.2$ for $L=4$ and about $0.3$ for $L=5$.
In Table~\ref{tab_gs_4D} we compare the average energy at $T=0.2$ with
the average ground state energy obtained with the hybrid genetic algorithm
of Ref.~\onlinecite{py2}. As in 3D, the data indicate that the system
is very close to the ground state.

\begin{table}
\begin{center}
\begin{tabular}{lrrr}
L  &  $\nsa^{(*)}$  & $\nsw$ & $N_T$  \\ 
\hline
3  & 60000 & $6 \times 10^3$ & 12    \\
4  & 30000 & $6 \times 10^4$ & 12   \\
5  & 12190 & $3 \times 10^5$ & 23    \\
\end{tabular}
\end{center}
\caption{Parameters of the simulations in four dimensions. $^{(*)}$ Quantities 
involving the link overlap $\ql$ have been calculated with
half the number of samples.
}
\label{4d-tab}
\end{table}
\begin{table}
\begin{center}
\begin{tabular}{lrrr}
$L$  &  $\langle E \rangle$  & $\langle E_0\rangle$ & $N_0$  \\ 
\hline
3 & $-158.64  \pm  0.03$  & $-158.82   \pm 0.04$   &  45000 \\
4 & $-510.42  \pm 0.07$ & $-510.93   \pm 0.08$ & 26681 \\
5 & $-1253.41 \pm 0.20$ & $-1254.33 \pm  0.20 $  &  8990
\end{tabular}
\end{center}
\caption{Average energy $\langle E \rangle$ at $T=0.2$ 
and average ground state energy $\langle E_0 \rangle$, for several
sizes in four dimensions. $N_0$ is the number of samples used to compute the
average $\langle E_0 \rangle$ using a hybrid genetic algorithm.}
\label{tab_gs_4D}
\end{table}

\begin{figure}
\begin{center}
\epsfxsize=\columnwidth
\epsfbox{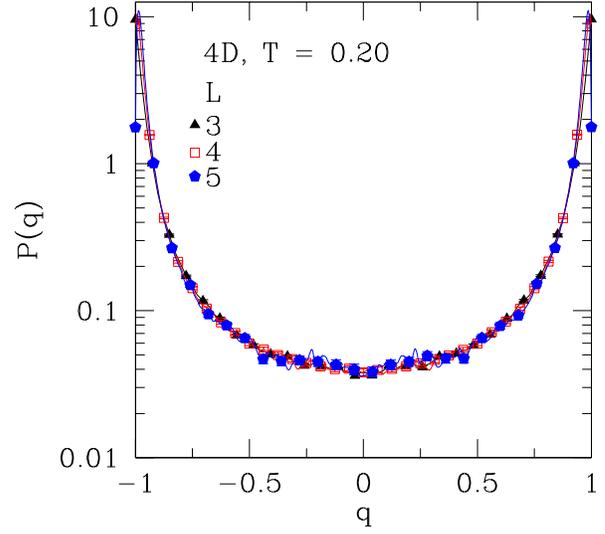}
\end{center}
\caption{
Data for the overlap distribution $P(q)$ in 4D at $T=0.20$.  The data is
normalized so the area under the curve is unity. Hence $P(q)$ is half as big
as it
it would be if we had just plotted the region of positive $q$ as in
Figs.~\ref{pq0.20_3d} and \ref{pq0.50_3d}.
}
\label{pq0.20_4d}
\end{figure}

\begin{figure}
\begin{center}
\epsfxsize=\columnwidth
\epsfbox{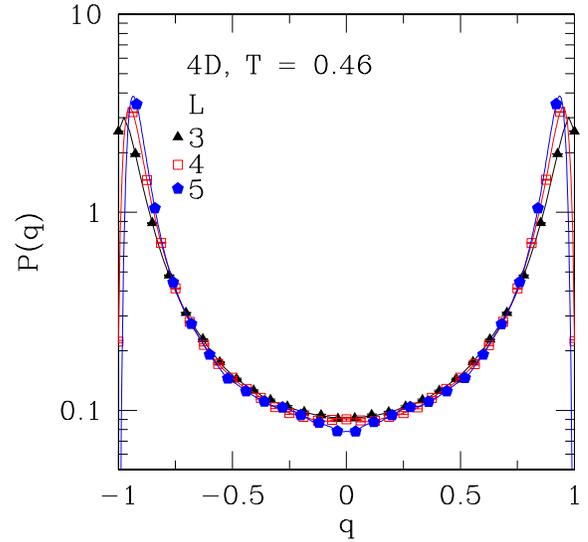}
\end{center}
\caption{
Same as for Fig.~\ref{pq0.20_4d} but at $T=0.46$.
}
\label{pq0.46_4d}
\end{figure}

Figs.~\ref{pq0.20_4d} and \ref{pq0.46_4d}
show data for $P(q)$ for temperatures 0.20
and 0.46.
As in three-dimensions, the tail in the distribution is
essentially independent of size. We display the full $P(q)$,
rather than just the symmetric part as in 3D, in order to show 
that it has a symmetric form as expected (a symmetric form was also 
obtained in 3D).

\begin{figure}
\begin{center}
\epsfxsize=\columnwidth
\epsfbox{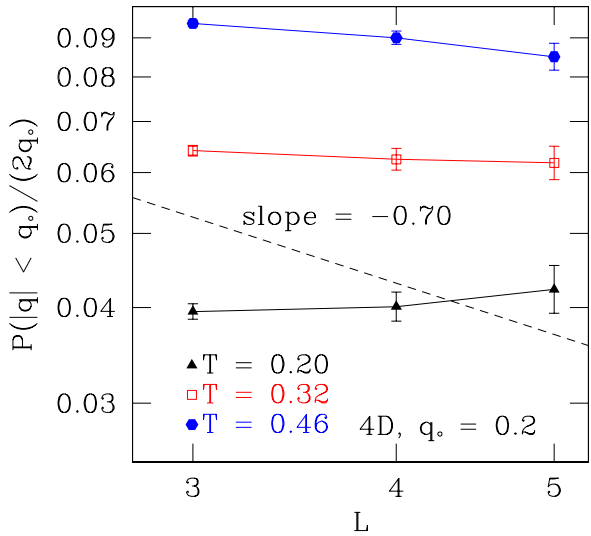}
\end{center}
\caption{
Log-log plot of $P(0)$ against $L$ in 4D averaged over the range $|q| < 0.20$.
The data is independent of size within the error bars.
The dashed line has slope $-0.70$, which is the estimated value of $-\theta$.
Asymptotically, the data should be parallel to this line according to the
droplet theory. 
}
\label{p0_4d}
\end{figure}

The size dependence of $P(0)$, averaged over the range $|q| < q_\circ$ with
$q_\circ = 0.2$ 
is shown in Fig.~\ref{p0_4d}.
The dashed line has slope $-0.70$
corresponding to the estimated value of $-\theta$.
In the droplet picture, the behavior should follow this form asymptotically.
Clearly it does not for this small range of sizes.
More precisely, performing a similar analysis as in 3D we find
$\theta^\prime =  0.10 \pm 0.12, 0.08 \pm 0.09$ and $0.17
\pm 0.06$ for $T=0.20, 0.32$ and $0.46$ respectively. For the
same temperatures, the goodness-of-fit parameter is
$0.67, 0.65$ and $0.014$,
assuming $\theta^\prime =0 $,
which is acceptable,
while assuming $\theta^\prime = 0.70 $ the goodness-of-fit parameters are
tiny: $2.1 \times 10^{-10} , 2.1 \times 10^{-12} $ and $ 4.2 \times 10^{-18}$.

As in 3D we find different results from what is predicted
by Refs.~\onlinecite{moore1,moore2}
at such low temperatures
on the basis of the the Migdal-Kadanoff
approximation.
However,
our data can not rule out the possibility that some other behavior may occur
at larger sizes.

\begin{figure}
\begin{center}
\epsfxsize=\columnwidth
\epsfbox{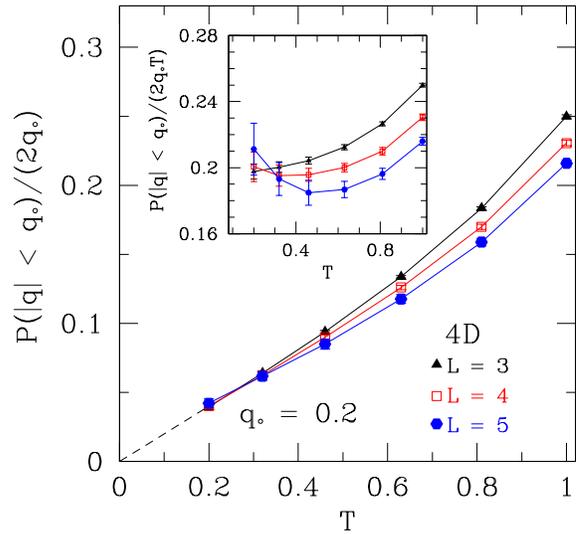}
\end{center}
\caption{
Data for $P(0)$ as a function of temperature in 4D for $L = 3,4,5$.
The inset shows $P(0)/T$ vs.~$T$.
}
\label{p0T_4d}
\end{figure}

Note also, that our data for $P(0)$ decreases approximately linearly with
temperature
as $T \rightarrow 0$, as shown in Figure \ref{p0T_4d}.

\begin{figure}
\begin{center}
\epsfxsize=\columnwidth
\epsfbox{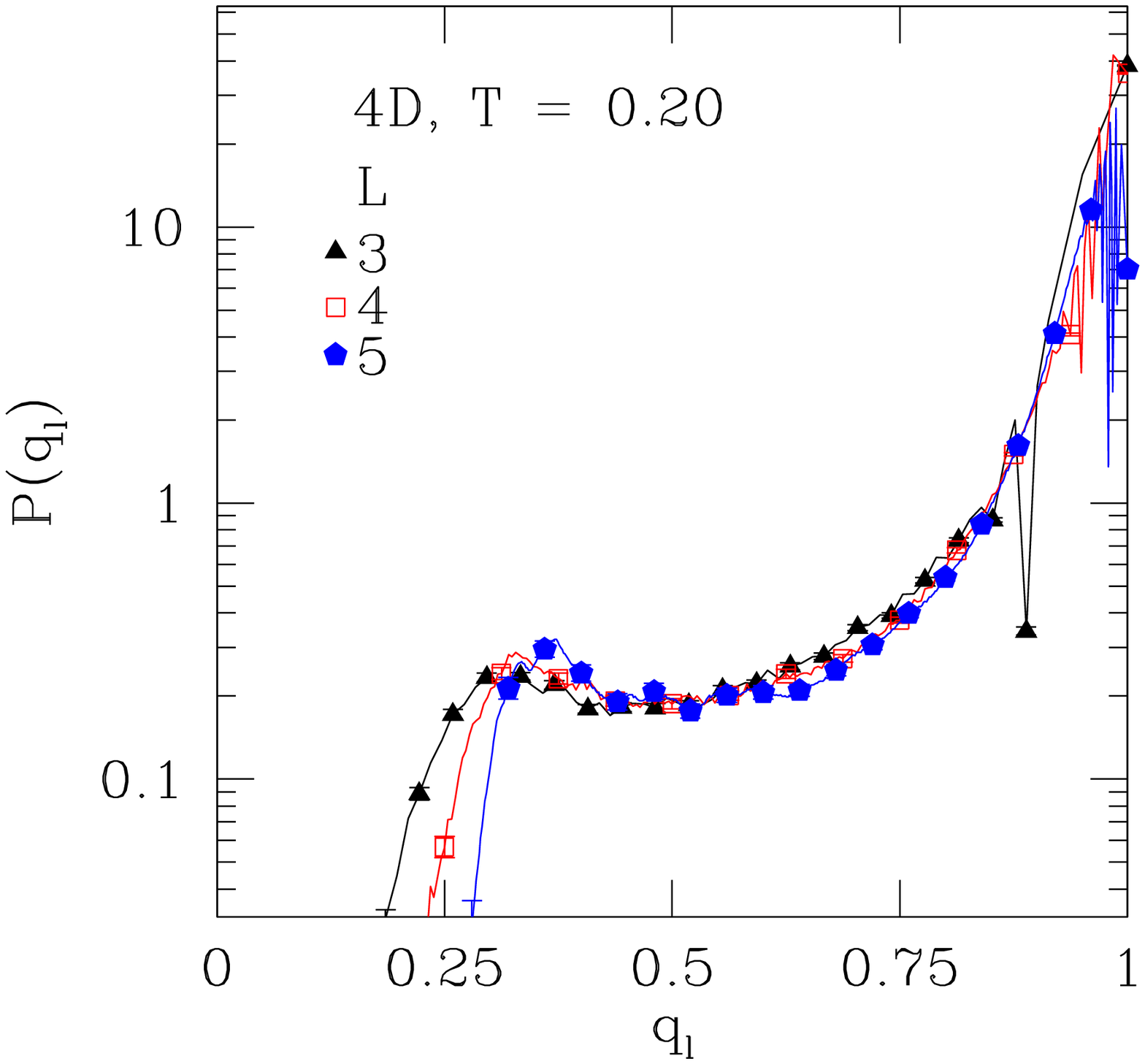}
\end{center}
\caption{
The distribution of the link overlap in 4D at $T=0.20$ for different sizes.
Note the logarithmic vertical scale.
}
\label{pqb0.20_4d}
\end{figure}

\begin{figure}
\begin{center}
\epsfxsize=\columnwidth
\epsfbox{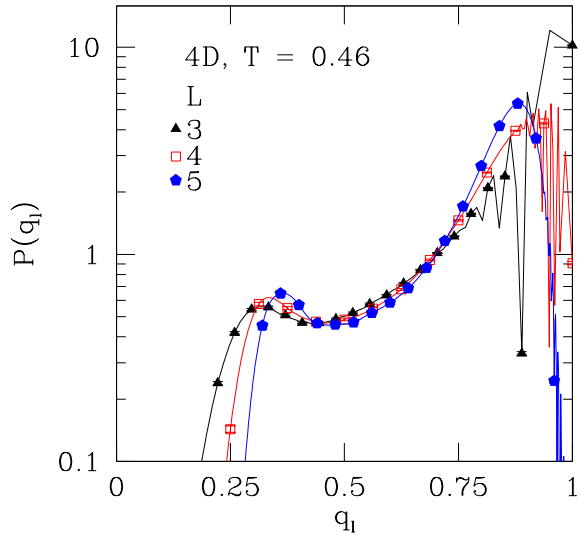}
\end{center}
\caption{
Same as for Fig.~\ref{pqb0.20_4d} but at temperature 0.46.
}
\label{pqb0.46_4d}

\end{figure}

Figs.~\ref{pqb0.20_4d} and \ref{pqb0.46_4d}
show the distribution of the link overlap
$\ql$ at temperatures 0.20 and 0.46.
As in 3D we see complicated structure at large $\ql$ and
a subsidiary peak at smaller $\ql$ which
grows with increasing $L$.

\begin{figure}
\begin{center}
\epsfxsize=\columnwidth
\epsfbox{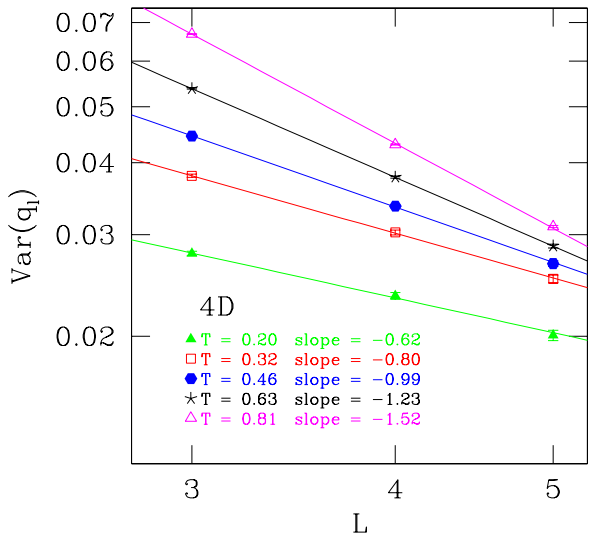}
\end{center}
\caption{
Log-log plot of the variance of $\ql$ as a function of size in 4D at several
temperatures. The data for $T = 0.63$ and $T = 0.81$ are multiplied by $1.15$
and $1.7$ respectively for better viewing. 
}
\label{varqb_4d}
\end{figure}

The variance of $\ql$ is shown in Fig.~\ref{varqb_4d} at
several low temperatures.
The data is consistent with the power law decrease to zero shown in
Eq.~(\ref{varqlpower}). The range of sizes is so small, and the
values of $\mu_l$ also so small, that we are not able to rule out a non-zero
value for $L \to \infty$ in 4D. However, the data is {\em consistent}\/ with
the asymptotic value being zero.

An extrapolation of our effective values of $\mu_l$ to $T=0$ gives $0.35 \pm
0.06$,  
which, assuming $\theta^\prime = 0$, gives 
\begin{equation}
d-d_s = 0.17 \pm 0.03.
\end{equation}

This is just consistent with the $T=0$ results of PY who
find $d - d_s = 0.21 \pm
0.01$. However, the quoted
error bars are from statistical errors only, so the
difference may be partly due to systematic effects coming from
the small range of sizes studied.

\subsection{Viana-Bray Model}
For the Viana-Bray model, $T_c$ is given by the solution of
\begin{equation}
{1 \over \sqrt{ 2 \pi} } \int_{-\infty}^\infty e^{-x^2/2} 
\tanh^2\left({x\over T_c}\right) \, dx = {1 \over z-1}, 
\end{equation}
where $z \ (=6\ \mathrm{here}) $ is the coordination number. The solution is
$T_c = 1.8075 \cdots $, which is roughly twice the transition
temperature of the 3D short range model considered here, which has 
the same coordination number.

Parameters of the simulations are shown in 
Table~\ref{vb-tab}. In each case, the largest temperature is 2.6 and the
lowest temperature is 0.1. For $N=700$ the data is not equilibrated for 
temperatures lower than $0.34$, and is almost equilibrated at $T=0.34$.
Except for $N=700$ the acceptance ratio for global moves is always greater
than about 0.3. For $N=700$ the acceptance ratio is greater than 0.3 for most
temperatures but there is one ``bottleneck'' where the acceptance ratio went
down to 0.08.

\begin{table}
\begin{center}
\begin{tabular}{lrrr}
N  &  $\nsa$  & $\nsw$ & $N_T$  \\ 
\hline
59   & 19022 & $ 10^4     $     &  21 \\
99   &  5326 & $  3\times10^4 $ &  21 \\
199  &	3116 & $ 10^5     $     &  21 \\
399  &	3320 & $ 10^5     $     &  21 \\
700  &	 801 & $ 3\times10^5 $  &  21
\end{tabular}
\end{center}
\caption{Parameters of the simulations for the Viana-Bray model.
}
\label{vb-tab}
\end{table}

\begin{figure}
\begin{center}
\epsfxsize=\columnwidth
\epsfbox{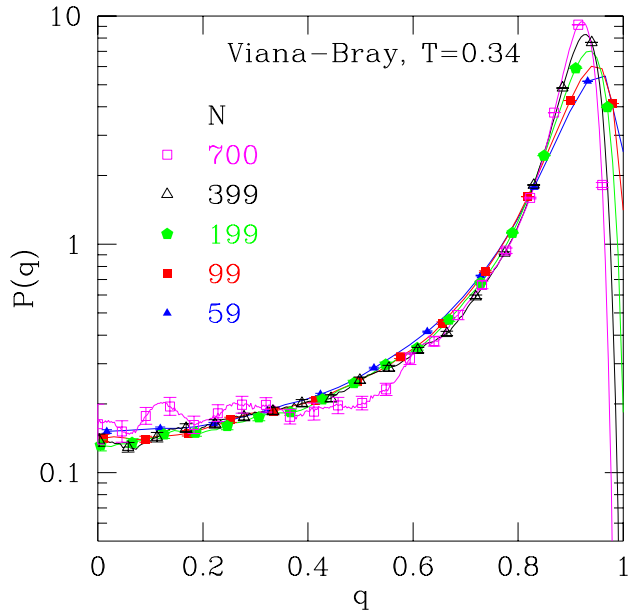}
\end{center}
\caption{
Data for the distribution of the overlap for the Viana-Bray model at
$T=0.34$. Note the logarithmic vertical scale.
}
\label{pq0.34_vb}
\end{figure}

First of all, in Fig.~\ref{pq0.34_vb} we show that $P(q)$ has a weight
at $q=0$ which appears to be independent of the system size, as expected.

\begin{figure}
\begin{center}
\epsfxsize=\columnwidth
\epsfbox{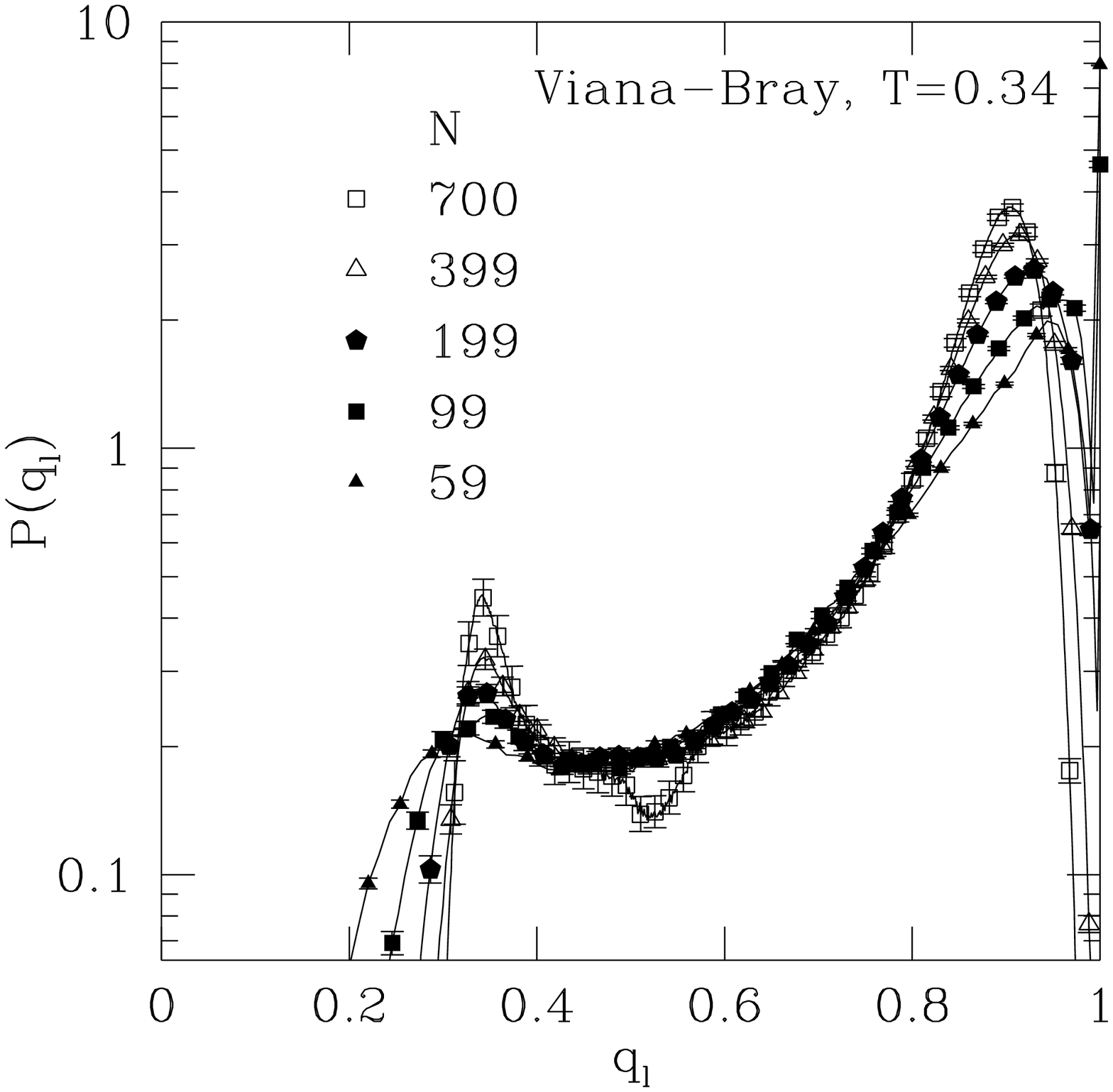}
\end{center}
\caption{
Data for the distribution of the link overlap for the Viana-Bray model at
$T=0.34$. Note the logarithmic vertical scale.
}
\label{pqb0.34_vb}
\end{figure}

\begin{figure}
\begin{center}
\epsfxsize=\columnwidth
\epsfbox{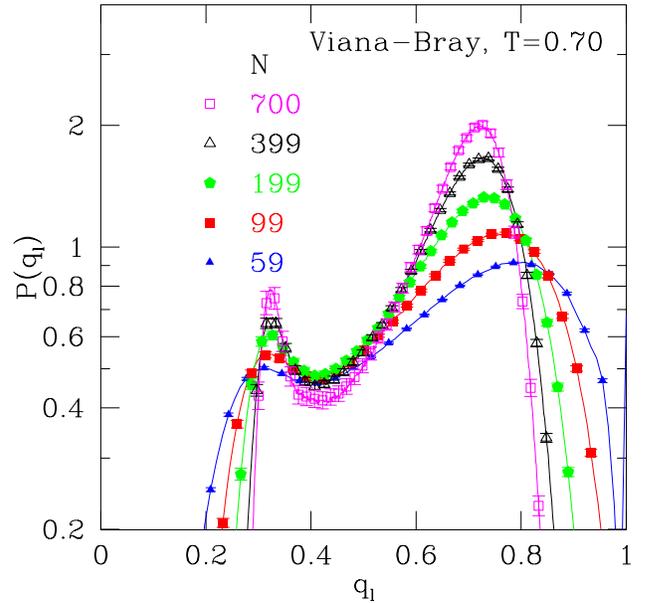}
\end{center}
\caption{
Same as for Fig.~\ref{pqb0.34_vb} but for $T=0.70$.
}
\label{pqb0.70_vb}
\end{figure}

A plot of $P(\ql)$ is shown in Figs.~\ref{pqb0.34_vb} and \ref{pqb0.70_vb}
for $T=0.34$ and $0.70$. Note that the data for $N=700$, $T=0.34$ show 
a dip around $q\simeq 0.5$, due to imperfect equilibration
of the Monte Carlo runs for some samples 
(this explains also the fluctuations of the
data for $P(q)$ in Fig.~\ref{pq0.34_vb}).
As in the 3D and 4D data
discussed above we see a 2-peak structure develop as the
size increases. This is also clearly visible in the earlier work of Ciria et
al.\cite{ciria} in 4D.
For the Viana-Bray model, the position of the smaller
peak shifts neither with $L$ or $T$ whereas for the 3D data, see
Fig.~\ref{pqb0.20_3d}, it clearly
shifts to larger values of $\ql$ with increasing size.  In 4D,
see Figs.~\ref{pqb0.20_4d} and \ref{pqb0.46_4d}, the range of sizes
is sufficiently small that it is difficult to tell whether there is a shift in
the position of the peak or not, but if it is present, it appears to be a
smaller effect than in 3D. It would be helpful to understand the physics
behind the two peak structure. For the hierarchical lattice used in the
Migdal-Kadanoff approximation, Bokil et al.\cite{moore2} have given an
explanation, but it is not clear to us how this goes over to the models
discussed here. 

\begin{figure}
\begin{center}
\epsfxsize=\columnwidth
\epsfbox{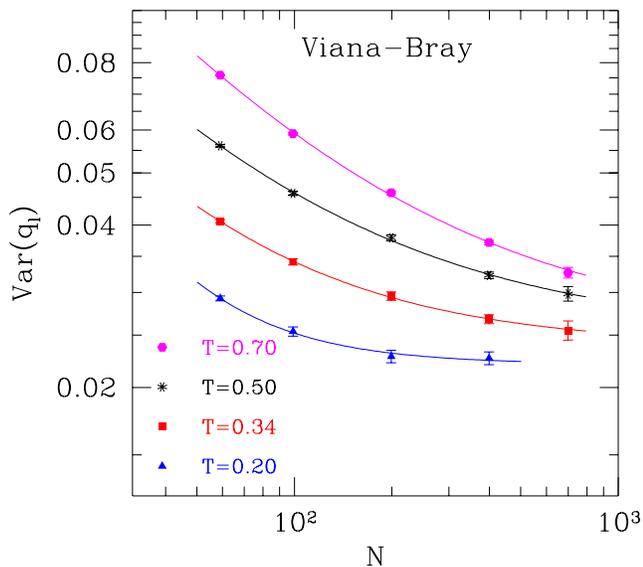}
\end{center}
\caption{
A log-log plot of the variance of $\ql$ for the Viana-Bray model at
different temperatures. The data at $T=0.5$ and $T=0.7$ are
multiplied by $1.2$ and $1.6$ respectively for better viewing.
}
\label{varqb_vb}
\end{figure}

A plot of $\varql$ against $N$
is shown in Fig.~\ref{varqb_vb} for different temperatures. 
In contrast to the data for 3D shown in Fig.~\ref{varqb_3d},
(which is for a similar number of spins)
the data is clearly tending to a
constant at large $N$.

\begin{figure}
\begin{center}
\epsfxsize=\columnwidth
\epsfbox{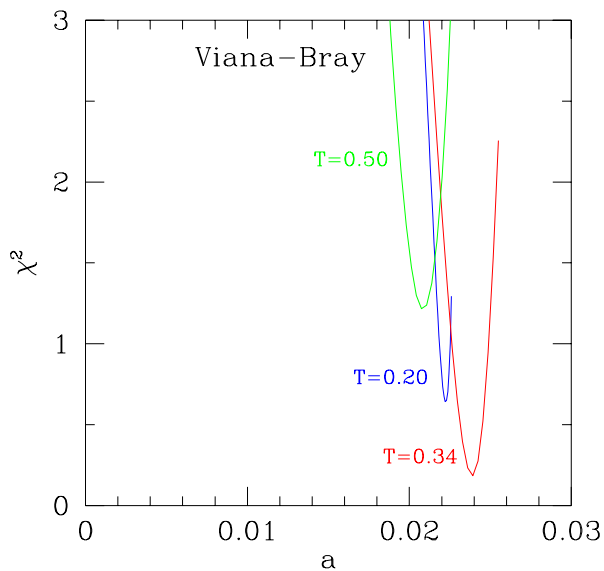}
\end{center}
\caption{
The $\chi^2$ of the fit in Eq.~(\ref{fit}) 
for different values of $a$ for the Viana-Bray model.
Note that the minima of the different $\chi^2$ are not monotonic in $T$
since we have fewer data points for $T = 0.20$.
}
\label{chi2_vb}
\end{figure}

This is confirmed by the $\chi^2$ analysis of the fits corresponding to
Eq.~(\ref{fit}) shown in Fig.~\ref{chi2_vb}. Clearly the
asymptotic value of $a$ is large and finite. Compare this figure with
Fig.~\ref{chi2_3d}, which shows the corresponding results in 3D, and where
$a=0$ gives the best fit. 

We conclude this section by pointing out the sizes studied (which covers a
similar range to that in 3D and 4D) {\em are}\/ sufficient to determine the
correct asymptotic behavior for the Viana-Bray model.

\section{Conclusions}
To conclude, Monte Carlo simulations at low (but finite) temperatures
agree with earlier $T=0$ studies of KM and PY
that there appear to be
large-scale low energy excitations which cost a finite energy, and
whose surface has fractal dimension
less than $d$.
However, since the sizes that we study are quite small,
there could be a crossover
at larger sizes to different behavior,
such as the droplet theory (with $\theta^\prime = \theta \ (> 0)$) or
an RSB picture (where $\theta^\prime = 0, d - d_s = 0$).
We note, however, that
our results for short range models
are quite different from those of the mean-field
like Viana-Bray model for samples with a 
similar number of spins, and, furthermore, our results for the Viana-Bray
model do predict the correct asymptotic behavior for that model.

\acknowledgements
We would like to thank
D.~S.~Fisher,
G.~Parisi, E. Marinari,
O.~Martin, M.~M\'ezard, 
M.~A.~Moore, H.~Bokil and
A.~J.~Bray for helpful discussions and correspondence.
This work was supported by the National Science Foundation under grants DMR
9713977 and 0086287.
The numerical calculations were made possible by a grant of time from
the National Partnership for Advanced Computational Infrastructure, and by use
of the UCSC Physics graduate computing cluster funded by the Department of
Education Graduate Assistance in Areas of National Need program.



\begin{references}

\bibitem{fh}
    D.~S.~Fisher and D.~A.~Huse, J. Phys. A. {\bf 20} L997 (1987); D.~A.~Huse
    and D.~S.~Fisher, J. Phys. A. {\bf 20} L1005 (1987); D.~S.~Fisher and
    D.~A.~Huse, Phys. Rev. B {\bf 38} 386 (1988).

\bibitem{bm}
    A.~J.~Bray and M.~A.~Moore, in {\em Heidelberg Colloquium on Glassy
    Dynamics and Optimization}\/, L.~Van~Hemmen and I.~Morgenstern eds.
    (Springer-Verlag, Heidelberg, 1986).

\bibitem{mcmillan}
    W.~L.~McMillan, J. Phys. C, {\bf 17}, 3179 (1984).

\bibitem{parisi}
    G.~Parisi, Phys. Rev. Lett. {\bf 43}, 1754 (1979); J. Phys. A {\bf 13},
    1101, 1887, L115 (1980; Phys. Rev. Lett. {\bf 50}, 1946 (1983).

\bibitem{mpv}
    M.~M\'ezard, G.~Parisi and M.~A.~Virasoro, {\em Spin Glass Theory and
    Beyond} (World Scientific, Singapore, 1987).

\bibitem{bindery}
    K.~Binder and A.~P.~Young, Rev. Mod. Phys. {\bf 58} 801 (1986).

\bibitem{ns}
    C.~M.~Newman and D.~L.~Stein, Phys. Rev. B {\bf 46}, 973 (1992); Phys. Rev.
    Lett., {\bf 76} 515 (1996);
    Phys. Rev. E {\bf 57} 1356 (1998).

\bibitem{qlink}
    E.~Marinari, G.~Parisi, F.~Ricci-Tersenghi, J.~Ruiz-Lorenzo and F.~Zuliani,
    J. Stat. Phys. {\bf 98}, 973 (2000).

\bibitem{km}
    F.~Krzakala and O.~C.~Martin, Phys. Rev. Lett. {\bf 85}, 3013 (2000) 
    (referred to as KM).

\bibitem{py4}
    M.~Palassini and A.~P.~Young, Phys. Rev. Lett. {\bf 85}, 3017 (2000)
    (referred to as PY).

\bibitem{mp}
    E. Marinari and G. Parisi, Phys. Rev. B {\bf 62}, 11677 (2000).

\bibitem{aam}
  A.~A.~Middleton, cond-mat/0007375.

\bibitem{hg}
    N.~Hatano and J.~E.~Gubernatis, cond-mat/0008115.

\bibitem{mprz}
    E.~Marinari, G.~Parisi, F.~Ricci-Tersenghi and F.~Zuliani,
    J. Phys. A {\bf 34}, 383 (2001).

\bibitem{pypmj}
    M.~Palassini and A.~P.~Young, cond-mat/0012161.

\bibitem{hed}
    G.~Hed, A.~K.~Hartmann, E.~Domany, cond-mat/0012451.

\bibitem{rby}
    J.~D.~Reger, R.~N.~Bhatt and A.~P.~Young, Phys. Rev. Lett. {\bf 64}, 1859
    (1990).

\bibitem{marinari}
    E.~Marinari, G.~Parisi, and J.~J.~Ruiz-Lorenzo, in {\em Spin Glasses and
    Random Fields}, edited by A.~P.~Young (World Scientific, Singapore, 1998),
    and references therein.

\bibitem{zuliani}
    E. Marinari and F. Zuliani, J. Phys. A {\bf 32}, 7447 (1999).

\bibitem{moore1}
    M.~A.~Moore, H.~Bokil and B.~Drossel, Phys. Rev. Lett. {\bf 81} 4252,
    (1998).

\bibitem{moore2}
    H.~Bokil, B.~Drossel and M.~Moore, Phys. Rev. B {\bf 62}, 946 (2000)

\bibitem{comment1}
    The basic argument of Refs.~\onlinecite{moore1,moore2} is that, at the
    critical point, $P(0)$ {\em increases}\/ with $L$ because the width of the
    distribution shrinks (tending to 0 for $L \to \infty$), whereas, by
    assumption, below $T_c$ $P(0)$ {\em decreases}\/ to zero for $L \to
    \infty$. Hence, for intermediate sizes and temperatures, these two effects
    could roughly cancel leading to a constant $P(0)$. However, in the first
    study of $P(q)$ below $T_c$, Ref.~\onlinecite{rby} found that, while
    $P(0)$ in 4D indeed increased with $L$ for $T$ just below $T_c$, at the
    lowest temperatures studied $P(0)$ initially {\em decreased} before
    leveling out, which already casts some doubt on the claim of
    Ref.~\onlinecite{moore1,moore2}.

\bibitem{huk_nem}
    K.~Hukushima and K.~Nemoto, 
    J. Phys. Soc. Japan {\bf 65}, 1604
    (1996).

\bibitem{marinari_buda}
    E.~Marinari, {\em Advances in Computer Simulation}, edited by J. Kert\'esz
    and Imre Kondor (Springer-Verlag, Berlin 1998), p. 50, (cond-mat/9612010).

\bibitem{by}
    R.~N.~Bhatt and A.~P.~Young, Phys. Rev. Lett. {\bf 54}, 924 (1985);
    {\em ibid.}\/, Phys. Rev. B {\bf 37}, 5606 (1988).

\bibitem{huku_sk}
    K.~Hukushima, H.~Takayama, and H.~Yoshino, J. Phys. Soc. Japan {\bf 67},
    12 (1998).

\bibitem{vb}
    L. Viana and A.~J. Bray, J. Phys. C, {\bf 18}, 3037 (1985).

\bibitem{theta-3d}
    A.~K.~Hartmann, Phys. Rev. E {\bf 59}, 84 (1999). 
    A.~J.~Bray and M.~A.~Moore, J. Phys. C, {\bf 17}, L463 (1984);
    W.~L.~McMillan, Phys. Rev. B {\bf 30}, 476 (1984).

\bibitem{theta-4d}
    A.~K.~Hartmann, Phys. Rev. E {\bf 60}, 5135 (1999);
    K.~Hukushima, Phys. Rev. E {\bf 60}, 3606 (1999).

\bibitem{bmq2}
    A.~J.~Bray and M.~A.~Moore, J. Phys. C, {\bf 13}, 419 (1980).

\bibitem{3d}
    E.~Marinari, G.~Parisi and J. J. Ruiz-Lorenzo, Phys. Rev. B, {\bf 58},
    14852 (1998).

\bibitem{py2}
    M.~Palassini and A.~P.~Young, Phys. Rev. Lett. {\bf 83}, 5126 (1999).

\bibitem{dbmb}
    B.~Drossel, H.~Bokil, M.~A.~Moore and A.~J.~Bray, Euro. Phys. J. {\bf B
    13}, 369 (2000).

\bibitem{4d}
    G.~Parisi, F.~Ricci-Tersenghi and J. J. Ruiz-Lorenzo, J. Phys. A, {\bf 29},
    7943 (1996).

\bibitem{ciria}
    J.~C.~Ciria, G.~Parisi, and F.~Ritort, J. Phys. A, {\bf 26}, 6731 (1993).

\end{references}
\end{document}